\documentclass[a4paper,
               keeplastbox,   
               ]{jacow}
\usepackage{pdfpages,multirow,ragged2e} %
\usepackage[colorlinks=false,hidelinks]{hyperref}
\usepackage[caption=false]{subfig}
\makeatletter%
	\ifboolexpr{bool{xetex}}
	 {\renewcommand{\Gin@extensions}{.pdf,%
	                    .png,.jpg,.bmp,.pict,.tif,.psd,.mac,.sga,.tga,.gif,%
	                    .eps,.ps,%
	                    }}{}
\makeatother

\ifboolexpr{bool{xetex} or bool{luatex}} 
 {}                                      
 {\usepackage[utf8]{inputenc}}           

\usepackage[USenglish]{babel}

\usepackage{subfig}
\usepackage{lipsum}

\ifboolexpr{bool{jacowbiblatex}}%
 {%
\PassOptionsToPackage{maxnames=1}{biblatex}
  \addbibresource{/Users/miriam/Documents/sync_server/paperwork/Lebenslauf_und_Publikationsliste/references_and_publications_bibfile/references_new_main_used_since2021.bib}
  \renewbibmacro*{doi+eprint+url}{%
\iftoggle{bbx:url} 
{\iffieldundef{doi}{\usebibmacro{url+urldate}}{}} 
{}%
\newunit\newblock 
\iftoggle{bbx:eprint} {\usebibmacro{eprint}} 
{}%
\newunit\newblock 
\iftoggle{bbx:doi} 
{\printfield{doi}} {}}

\usepackage{biblatex2bibitem}
 }{}

\begin{document}

\title{Overview of the Micro-bunching Instability in Electron Storage Rings and Evolving Diagnostics}

\author{M. Brosi\thanks{miriam.brosi@kit.edu}, 
Karlsruhe Institute of Technology, Karlsruhe, Germany}
	
\maketitle

\begin{abstract}
The micro-bunching instability is a longitudinal instability that leads to dynamical deformations of the charge distribution in the longitudinal phase space. It affects the longitudinal charge distribution, and thus the emitted coherent synchrotron radiation spectra, as well as the energy distribution of the electron bunch. Not only the threshold in the bunch current above which the instability occurs, but also the dynamics above the instability threshold strongly depends on machine parameters, e.g. accelerating voltage, momentum compaction factor, and beam energy.
All this makes the understanding and potential mitigation or control of the micro-bunching instability an important topic for the next generation of light sources and circular e+/e- colliders.

This contribution will give an overview of the micro-bunching instability and discuss how technological advances in the turn-by-turn and bunch-by-bunch diagnostics are leading to a deeper understanding of this intriguing phenomenon.
   \end{abstract}
 
 \vspace{-0.2cm}
\section{Introduction}
The micro-bunching instability is a longitudinal, collective, single-bunch instability. 
It occurs during the operation of storage rings with short electron bunches above a certain threshold current. 
Although the instability can also be observed at linear accelerators, this contribution will focus on the instability in electron storage rings.

While not leading to instant beam loss, the micro-bunching instability leads to dynamic changes and deformations of the charge density in the longitudinal phase space deteriorating the beam properties. 
It therefore limits the bunch current range for stable operation with short electron bunches. 
The instability was observed and studied at many different electron storage rings in the world, amongst others at ALS~\cite{ALS2002}%
, ANKA~\cite{anke2005}%
, BESSY II~\cite{bakr2003}%
, CLS~\cite{Billinghurst_prab16}%
, DIAMOND~\cite{Shields_2012}%
, Elettra~\cite{KARANTZOULIS2010300}%
, MAX-I~\cite{andersson_1999}%
, MLS~\cite{mls_ipac10}%
, NewSUBARU~\cite{newsubaru_2005}%
, NSLS VUV Ring~\cite{carr1999}%
, UVSOR-II~\cite{Takashima_2005}%
, SLC damping ring~\cite{Podobedov:336088}%
, SOLEIL~\cite{soleil_2012} %
and SURF III~\cite{PhysRevSTAB.4.054401}.%

In the following, a short overview of the micro-bunching instability is given. 
First the underlying dynamics is introduced followed by a short outline of the theoretical description as well as simulation methods. Then some examples of different measurements conducted at different facilities are given. Afterwards, the focus is on the fast diagnostics developed at KIT and the measurements conducted with them. At the end some examples of ongoing studies concerning the micro-bunching instability are given.

\vspace{-0.15cm}
\section{micro-bunching instability}
\vspace{-0.15cm}
		
The micro-bunching instability occurs due to the self-interaction of an electron bunch with its own emitted coherent synchrotron radiation (CSR). 
Coherent emission occurs when the emitting structure is shorter than the emitted wavelength. 
The waveguide cut-off of the vacuum pipe suppresses the emission of longer wavelength depending on the dimensions of the vacuum pipe. 
For a bunch shorter than the waveguide cut-off, a wake potential due to the emitted CSR acts back on the bunch. 
Due to the bent path of the electrons in a bending magnet the emitted CSR can lead to a forward interaction. 
This self-interaction causes a change in the energy distribution which is transformed to a spacial change via the synchrotron motion. 
Due to the now deformed longitudinal bunch profile the emitted CSR spectrum changes, leading directly to a change of the wake potential and therefore the self-interaction. 

For bunch currents above the instability threshold, this continuous self-interaction leads to the formation of substructures in the charge distribution in the longitudinal phase-space. 
Depending on the exact parameters, the dynamics differ. 
For bunch currents close to the threshold current, the substructures are mostly stable in amplitude and rotate in the phase space due to synchrotron motion, leading to fast, repetitious changes in the emitted CSR. 
For higher bunch currents an additional slower dynamic occurs. 
The substructures start to increase in amplitude and are driven further by the also increasing wake potential until filamentation sets in. 
As the substructures wash out the driving wake potential becomes weaker and the previously blown up phase-space distribution starts to shrink and the bunch length is damped down again. 
This cycle starts anew as soon as the bunch is short enough for the wake potential to increase and to drive new substructures. 
The rising and damping of substructures leads to intense bursts in the emitted CSR with a low repetition rate. In parallel the bunch length and energy spread increase and decrease. 

These dynamics drastically influence the beam quality in the longitudinal as well as the horizontal plane as a change in the energy spread is coupled to a change in the horizontal size via the dispersion. 
The instability therefore limits the stable operation for high bunch currents with short electron bunches. 

\vspace{-0.2cm}
\section{theory \& simulation}
\vspace{-0.1cm}
The temporal development of the charge distribution $\psi$ in the longitudinal phase space $\left(q,p\right)$ can be described by the Vlasov-Fokker-Planck equation~\cite{Ng_book2006}:
\vspace{-0.2cm}
$$\frac{\mathrm{d}\psi}{\mathrm{d}\theta}=\frac{\partial\psi}{\partial\theta}+\frac{\partial\mathcal{H}}{\partial p}\frac{\partial\psi}{\partial q}-\frac{\partial\mathcal{H}}{\partial q}\frac{\partial\psi}{\partial p}=\beta_{\mathrm{d}}\frac{\partial}{\partial p}\left(p\psi+\frac{\partial\psi}{\partial p}\right).$$
The influence of the CSR self-interaction can be included as an additional, collective term in the Hamiltonian:
\vspace{-0.2cm}
\begin{gather*}
\mathcal{H}\left(q,p,\theta\right)=\underbrace{\mathcal{H}_{l}\left(q,p,\theta\right)}_{\mathrm{unperturbed}}+\underbrace{\mathcal{H}_{c}\left(q,\theta\right)}_{\mathrm{collective}}\\
=\frac{1}{2}\left(q^{2}+p^{2}\right)+\frac{ef_{\mathrm{rev}}}{\sigma_{E}f_{\mathrm{s,0}}}\int_{q}^{\infty}\!\!\!\int_{-\infty}^{\infty}\!\varrho\left(f\right)Z\left(f\right)e^{i2\pi fq'}\mathrm{d} f\mathrm{d} q'.
\vspace{-0.2cm}
\end{gather*}

For the CSR self-interaction leading to the micro-bunching instability the impedance $Z\left(f\right)$ is dominated by the CSR impedance $Z_{\mathrm{CSR}}\left(f\right)$. 
Different models for $Z_{\mathrm{CSR}}\left(f\right)$ exist. 
For most simulations either the parallel plates (shielded) model (e.g.~\cite{Warnock_1990,PhysRevSTAB.12.094402}) or the free-space (no shielding) model (e.g.~\cite{murphy1997}) are used. 
Both models assume an electron bunch flying on a circular path in vacuum. 
For the parallel plates model, two infinitely wide, perfectly conducting, parallel plates are added above and below the bunch path. 
These plates model the waveguide cut-off caused by the vacuum chamber. 
More sophisticated models exist, e.g.~\cite{PhysRevSTAB.12.094402}.

\begin{figure}
	\centering
	\includegraphics[width=1\columnwidth]{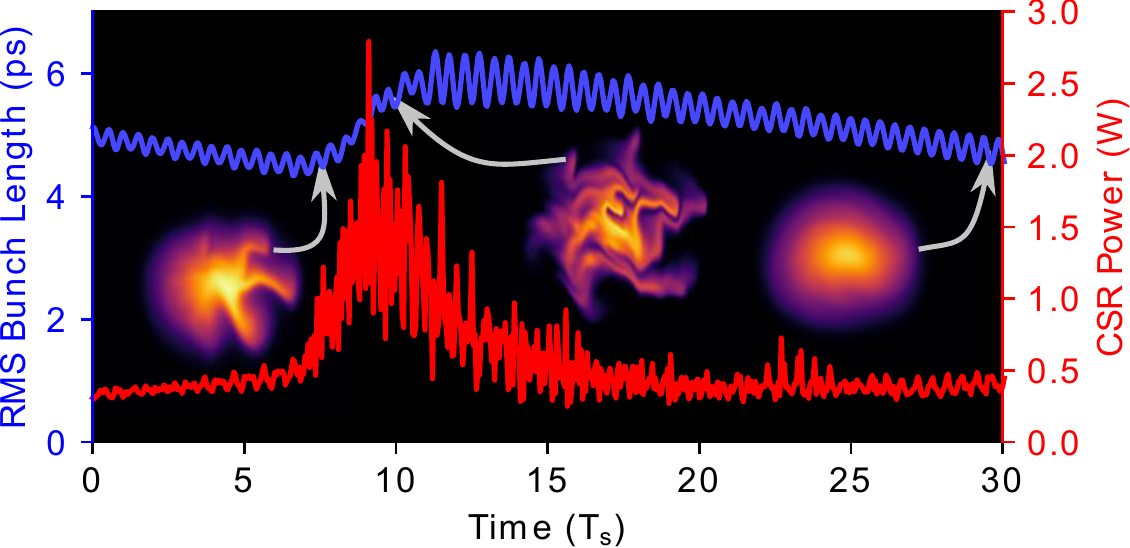}
	\caption{Simulation with VFP solver Inovesa~\cite{inovesa_github_2018} showing the development of the phase space distribution and the emitted CSR power as well as the bunch length. The substructures increase and are then washed out which results in a burst in the emitted CSR and a change in the bunch length. Courtesy of P. Schönfeldt.}
	\label{fig:sim}
			\vspace{-0.55cm}
\end{figure}

The micro-bunching instability can be simulated by solving the VFP equation iteratively in time and thereby observing the dynamics in the longitudinal phase space. 
This was for example done in~\cite{Podobedov:336088,warnock_2000,Stupakov2002,venturini2002,Bane_cai_stupakov2010}. 
Such a simulation can be seen in Fig.~\ref{fig:sim}. Here, the VFP solver Inovesa~\cite{inovesa_github_2018} was used to simulate the temporal development of the longitudinal phase space over 30 synchrotron periods. The growth and decrease of the substructures is visible as well as the burst in emitted CSR power and the change in bunch length. 

\vspace{-0.1cm}
\section{Measurement examples}

These dynamics of CSR bursts accompanied by changes in the bunch length as well as in the energy spread were observed, for example, in the early 2000s at the SURF III facility~\cite{PhysRevSTAB.4.054401}. 
As the instability was observed at many facility, there is a multitude of different measurements of the micro-bunching instability conducted at electron storage rings around the world. 
For example, a measurement of the average frequency spectrum of the emitted CSR during a burst was conducted at the ALS~\cite{ALS2002}.  
Or at the SLC damping rings, the emitted CSR was measured as a function of time showing how the power fluctuations change with bunch current~\cite{Podobedov:336088}. 
This is also nicely visible in the spectrograms of the CSR power fluctuations as a function of the bunch current recorded at the Diamond light source~\cite{Shields_2012}. 
At Soleil the substructures occurring on the bunch profile where observed by measuring the CSR pulse emitted by an electron bunch with a combination of electro-optic sampling and photonic time-stretch~\cite{soleil_far-fiel_2015}. 
The systematic dependence of the instability threshold from different operational parameters was studied, amongst others, at the MLS~\cite{ries_ipac12}. 
Measurements showing steps in the initial unstable mode (the instability frequency directly above the threshold) as a function of different operational settings resulting in different synchrotron frequencies were for example investigated at the Canadian Light Source~\cite{Billinghurst_prab16}. 

\vspace{-0.1cm}
\section{Micro-bunching at KARA}
KARA (KArlsruhe Research Accelerator) is the storage ring of the KIT light source in Karlsruhe, Germany. 
It is a ramping machine with a circumference of 110.4\,m and an RF frequency of 500\,MHz. 
Besides the standard low-emittance operation mode at 2.5\,GeV, KARA also provides short-bunch operation modes with a reduced momentum compaction factor $\alpha_\mathrm{c}$ at different beam energies as well as a negative momentum compaction operation mode~\cite{huttle2005,anke2005,Papash_ipac19,schreiber_ipac19}.  
This section will focus on fast, turn-by-turn and bunch-by-bunch diagnostic systems developed at KIT and show some example studies enabled by the technological advances. 

\vspace{-0.1cm}
\subsection{Fast Diagnostic Systems}

\vspace{-0.1cm}
\subsubsection{CSR Power} To measure the changes in the emitted CSR power, fast THz detectors are combined with the dedicated DAQ system KAPTURE~\cite{caselle_ibic14}. 
Mainly quasi-optical, broadband or waveguide-coupled, narrowband zero-biased Schottky Barrier diode detectors~\cite{vdi,acst}, all with an analog bandwidth $>$4\,GHz, are used as THz detectors.   
KAPTURE is short for KArlsruhe Pulse Taking Ultra-fast Readout Electronics and consists of four 500\,MS/s sampling channels (eight 1\,GHz channels for KAPTURE2~\cite{KAPTURE2_2017}) with a 12-bit ADC each. 
The delay between the sampling channels can be adjusted in 3\,ps steps achieving a local sampling rate of up to 330\,GS/s. 
This design allows the continuous sampling of the detector pulses for each bunch at every turn while the time between the pulses is not sampled reducing the data rate to 32\,Gb/s.  

\vspace{-0.25cm}
\subsubsection{Longitudinal Bunch Profile}
The longitudinal bunch profile is measured with a near-field electro-optical spectral decoding setup~\cite{hiller_diss} combined with a grating and the fast line array detector KALYPSO as spectrometer~\cite{KALYPSO_ibic16}. 
KALYPSO exists in different configurations, e.g. with 512, 1024 or 2048 pixels and different sensor materials~\cite{Patil_ipac19,Patil_ipac21}. 
For 512 pixels it provides a sampling rate of up to 10\,Mfps which allows turn-by-turn measurements in single-bunch operation.  
This combination allows turn-by-turn measurements of single-shot longitudinal bunch profiles.

\vspace{-0.25cm}
\subsubsection{Energy Spread}
The energy spread is connected to the horizontal bunch size via the dispersion. This connection is used by measuring the horizontal bunch size at a dispersive position (in this case a 5° port of a bending magnet) with source point imaging using light in the visible range~\cite{kehrer_ipac17}. The source image of the bunch is stretched in the horizontal plane to achieve a better resolution before it is recorded with KALYPSO~\cite{kehrer_ipac19}. KALYPSO is used here as fast line array detector which provides turn-by-turn readout.

\vspace{-0.15cm}
\subsection{Measurements}
With the diagnostic systems described above a multitude of measurements were conducted at KARA. The following section will give a short overview of some examples, most of them conducted at a beam energy of 1.3\,GeV. More measurements and studies can be found for example in~\cite{schoenfeldt_diss_18,steinmann_diss_18,Kehrer_diss2019,Brosi_thesis_2020}. 

\vspace{-0.25cm}
\subsubsection{Synchronized Measurements}
\begin{figure}
	\centering
	\includegraphics[width=1\columnwidth]{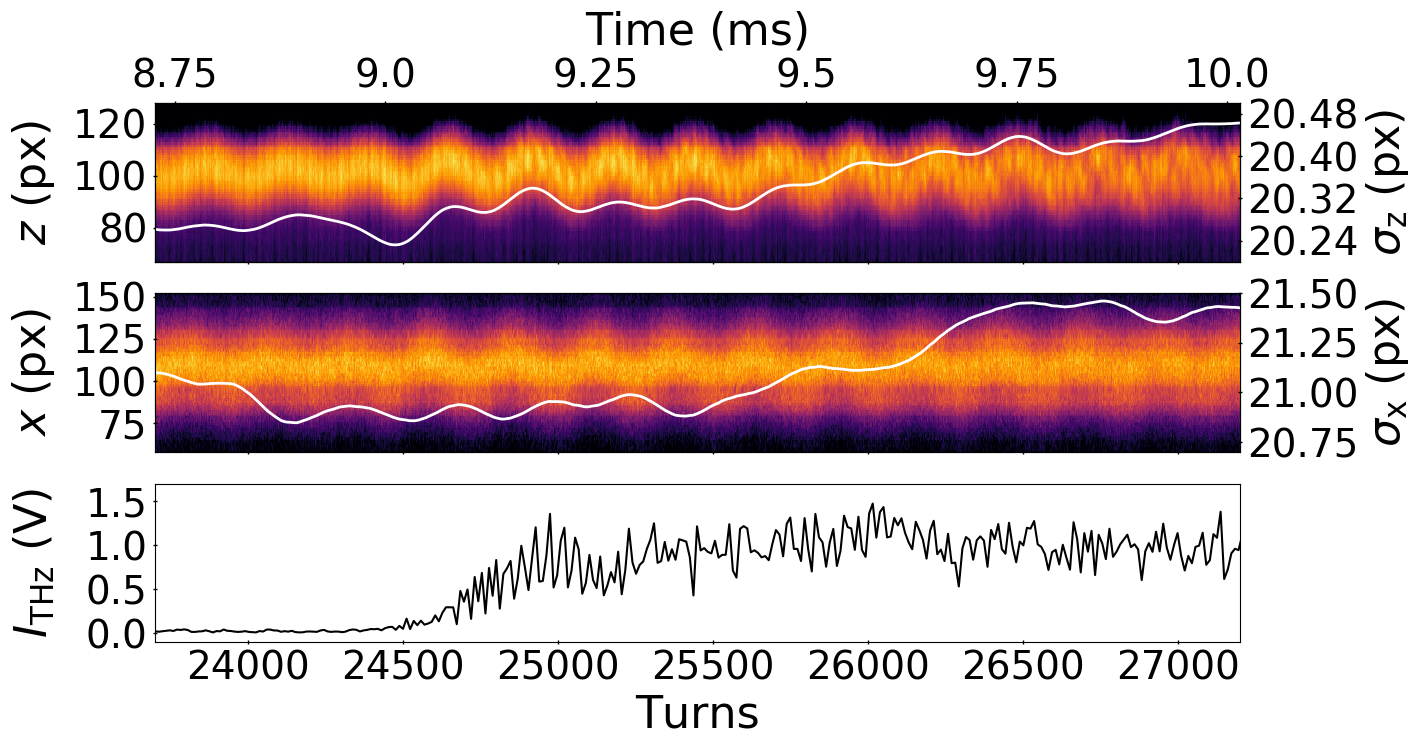}
	\vspace{-0.6cm}
	\caption{Synchronized measurement. From top to bottom: longitudinal profile with long. size ($\sigma_\mathrm{z}$) in white, horizontal profile with hor. size ($\sigma_\mathrm{x}$ proportional to energy spread) in white and THz detector signal ($I_\mathrm{THz}$).~\cite{Brosi_ipac19}}
	\label{fig:synchro}
			\vspace{-0.35cm}
\end{figure}

The diagnostic systems for the CSR power, the longitudinal bunch profile and the energy spread can be synchronized with a common acquisition trigger on a turn basis~\cite{kehrer_prab18}. 
Such a measurement is shown in Fig.~\ref{fig:synchro}.
The first observation is that at the same time the emitted CSR power increases substructures become visible on the longitudinal bunch profile.
Similar measurements showing the formation of substructures were conducted at Soleil with a far-field electro-optical setup~\cite{soleil_far-fiel_2015}.  
Contrary to the observations in simulations, no substructures are visible on the horizontal profile, which is attributed to a limited spatial resolution~\cite{Brosi_ipac19}. 
Nevertheless, the expected increase in bunch length as well as horizontal size (corresponding to the energy spread) during the burst can be observed similar to other measurements e.g. at SURF III~\cite{PhysRevSTAB.4.054401}.
Due to the on-turn synchronization, a small delay in the increase of the horizontal size could be observed. 
Further measurements showing several burst can be found in~\cite{Kehrer_diss2019,Brosi_ipac19}. 
Additionally, small changes in the longitudinal position (not shown here) could be observed. 
The amplitude of the shift corresponds to a phase change of the bunch to compensate the additional energy loss due to the increased CSR emission~\cite{Brosi_ipac19}. 

\vspace{-0.3cm}
\subsubsection{Influence of Operational Parameters}
\begin{figure}
	\centering
	\includegraphics[width=1\columnwidth]{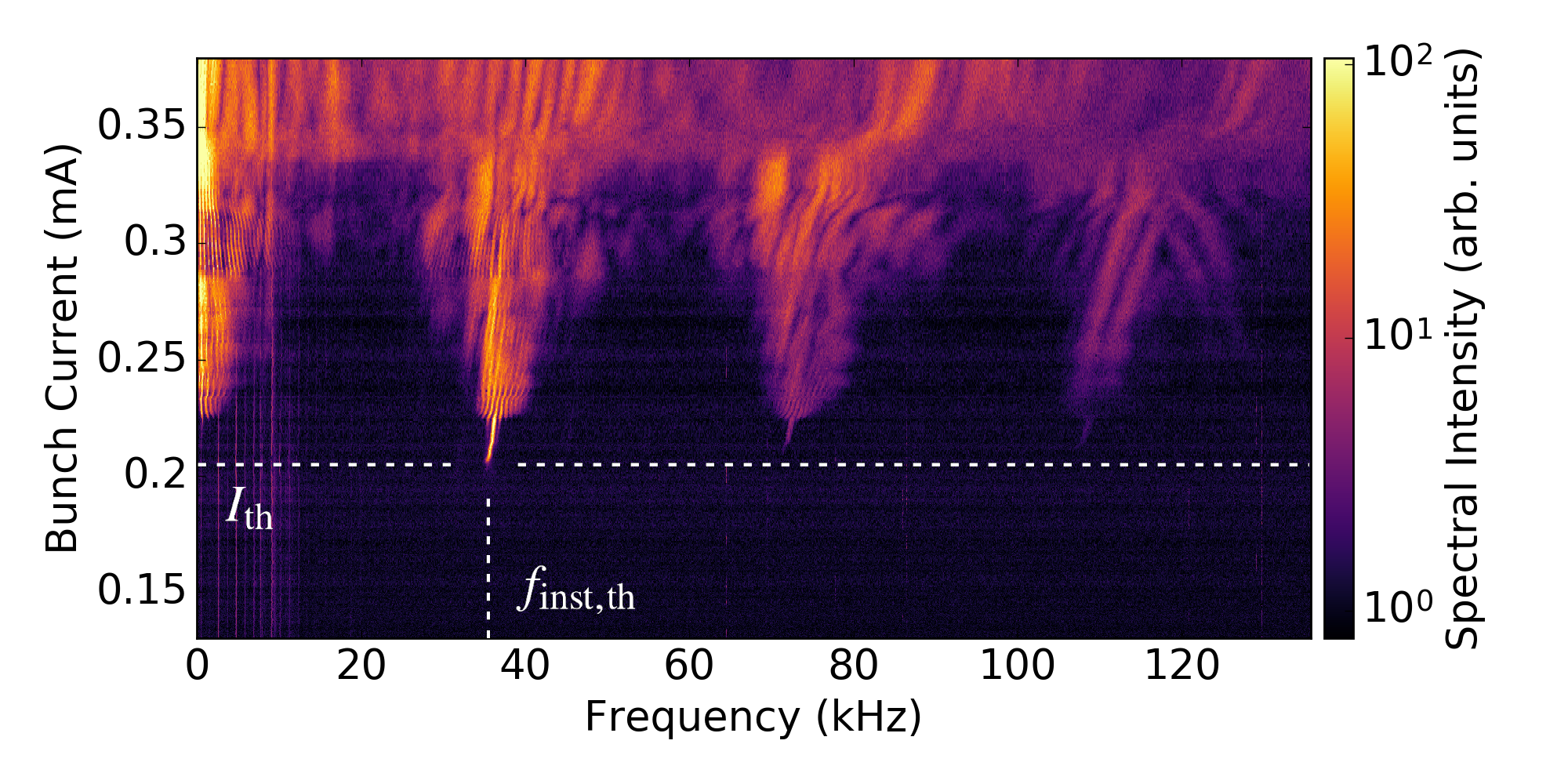}
	\vspace{-0.7cm}
	\caption{Spectrogram showing the fluctuation frequencies in the measured CSR power in the THz frequency range as a function of bunch current. Modified from~\cite{brosi_prab16}.}
	\label{fig:spec}
			\vspace{-0.55cm}
\end{figure}

The emitted CSR power is a good indicator for the underlying dynamics of the instability, with the benefit that it can also be measured for each bunch in multi-bunch operation by using KAPTURE. 
In the following the main characteristics of the micro-bunching instability are described and studies of their dependence on operational parameters are shortly summarized.
Figure~\ref{fig:spec} shows a typical measurement, where the emitted CSR power of one bunch was measured over several thousand turns for different bunch currents. 
Displayed is the Fourier transform of the time domain CSR signal for each bunch current, showing the fluctuation frequencies in the CSR emission. 
Such a spectrogram is a reproducible fingerprint, from which the main characteristics can be extracted~\cite{brosi_prab16}.
The instability threshold $I_\mathrm{th}$ is the bunch current below which no fluctuations due to the instability are visible. 
And the initial unstable mode also referred to as the instability frequency $f_\mathrm{inst,th}$ directly above the threshold current is visible as the lower end of the finger-like structure. 
The third characteristic is the repetition rate of the characteristic bursts. 
In Fig.~\ref{fig:spec} it is only visible as a bright area at the left edge of the plot as it is at frequencies of only a few hundred Hertz, therefore it is also referred to as low bursting frequency. 

Based on these three characteristic properties, the influence of different operational parameters was studied. 
For example, it could be shown, that a change in the longitudinal damping time (for otherwise unchanged parameters) leads to a change of the low bursting frequency but did not influence $I_\mathrm{th}$ and $f_\mathrm{inst,th}$~\cite{brosi_ipac18}. 
The damping time was changed using a CLIC damping ring wiggler prototype installed at KARA~\cite{Bernhard_ipac16}. 
Recent experiments showed, that the beam energy has a similar influence on the low bursting frequency but, as expected, also influences $I_\mathrm{th}$ and $f_\mathrm{inst,th}$~\cite{Brosi_ipac21_contributed}.

\begin{figure}
\centering
	\includegraphics[width=0.499\columnwidth]{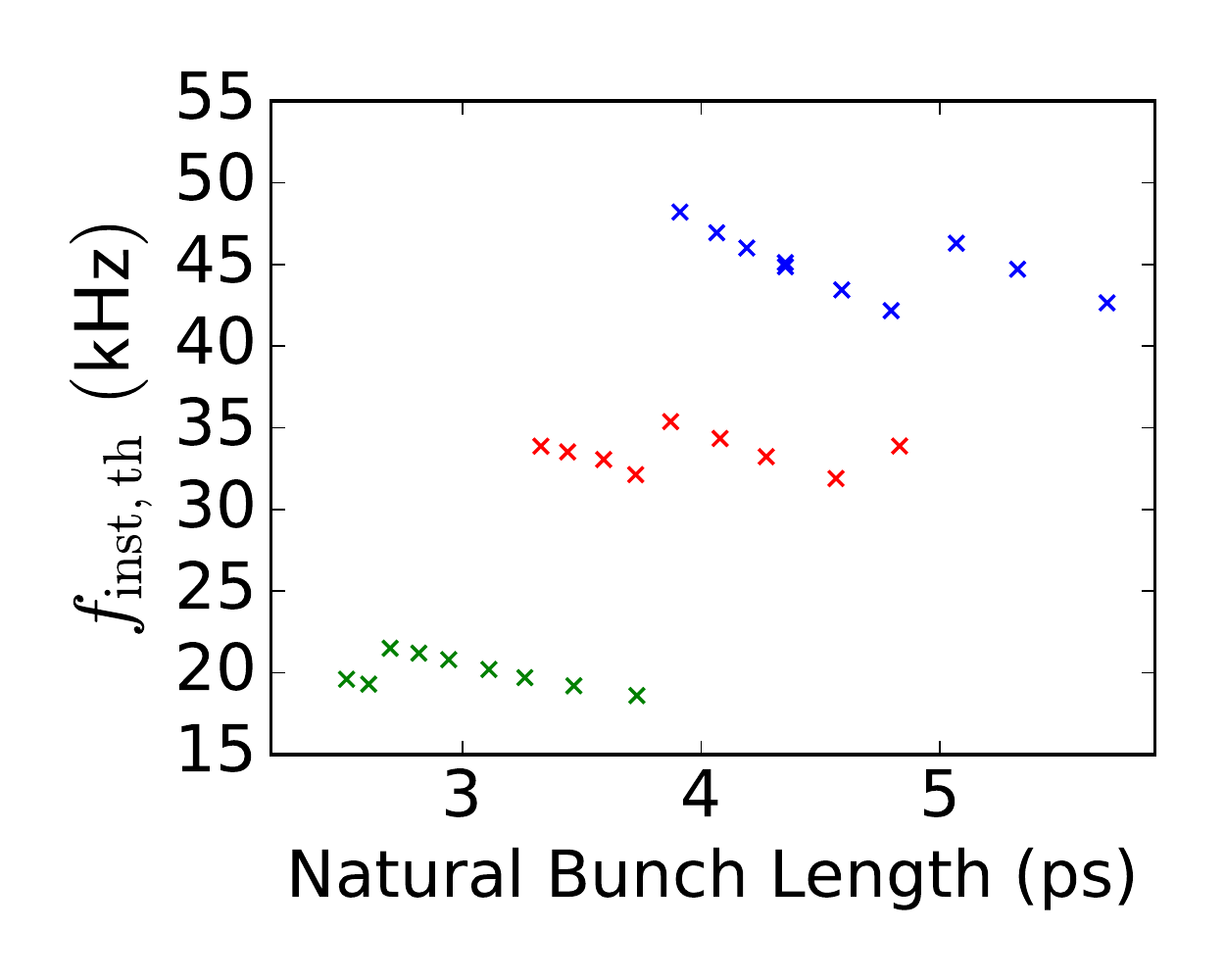}
	\hspace{-0.2cm}\includegraphics[width=0.499\columnwidth]{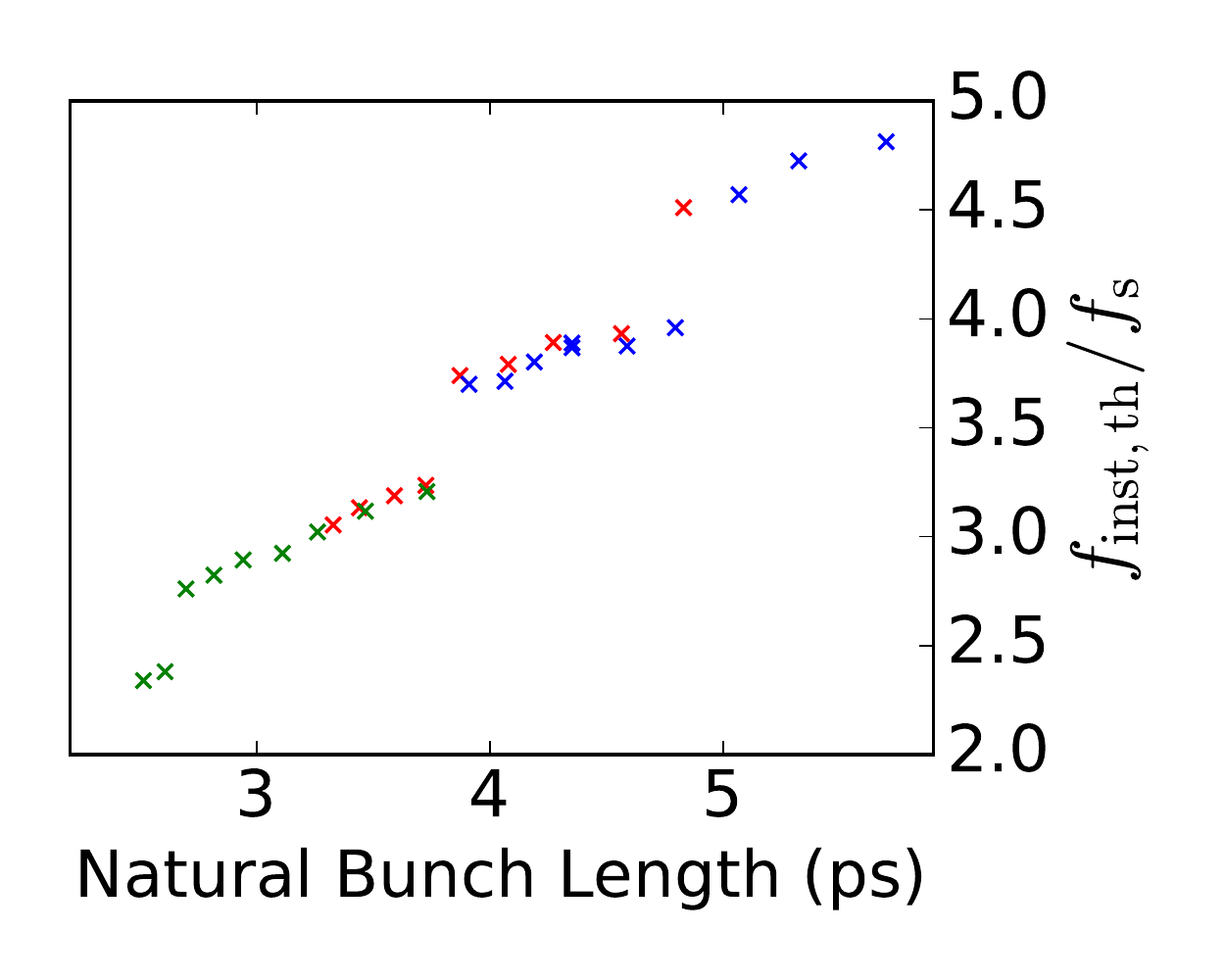}

\vspace{-0.25cm}
	\includegraphics[width=1\columnwidth,trim=10 0 0 0,clip]{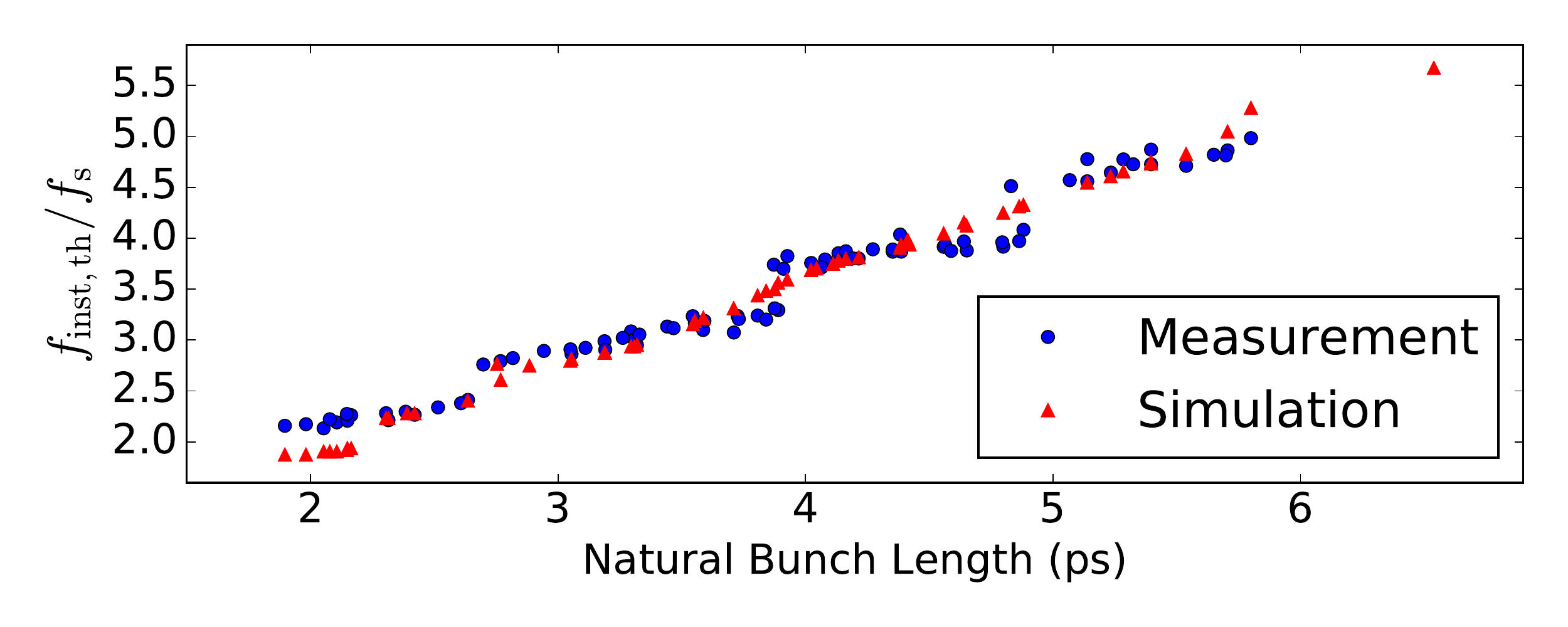}
	\vspace{-0.75cm}
	\caption{Initial unstable mode ($\hat{=}$ instability frequency at threshold) $f_\mathrm{inst,th}$ as a function of natural bunch length $\sigma_\mathrm{z,0}$. Displayed as $f_\mathrm{inst,th}/f_\mathrm{s}$ the connection to $\sigma_\mathrm{z,0}$ becomes unambiguous. In simulations (courtesy of P. Kuske, HZB,~\cite{Kuske_ipac17}) the steps are not as pronounced as in the measurements.~\cite{Brosi_thesis_2020}}
	\label{fig:finger}
			\vspace{-0.3cm}
\end{figure}

Other operational parameters, like the momentum compaction factor $\alpha_\mathrm{c}$  and the acceleration voltage $V_\mathrm{RF}$ have an extensive influence on $I_\mathrm{th}$ and $f_\mathrm{inst,th}$ as they both influence the bunch length and the restoring force in the longitudinal direction. 
Figure~\ref{fig:finger} shows the bursting frequency directly above the threshold $f_\mathrm{inst,th}$ for different measurements at different machine settings resulting in different natural bunch lengths $\sigma_\mathrm{z,0}$.
Note, that the same value of $\sigma_\mathrm{z,0}$ can be reached with different combinations of $\alpha_\mathrm{c}$ and $V_\mathrm{RF}$. 
This results in the possibility to have different values of $f_\mathrm{inst,th}$ for a certain $\sigma_\mathrm{z,0}$ as can be seen in the upper left panel of Fig.~\ref{fig:finger}. 
When displaying the bursting frequency in multiples of the corresponding synchrotron frequency ($f_\mathrm{inst,th}/f_\mathrm{s}$) the different values collapse to one~\cite{Brosi_thesis_2020}. 
The quotient $f_\mathrm{inst,th}/f_\mathrm{s}$ is then unambiguous for a given value of $\sigma_\mathrm{z,0}$.
Like in measurements at CLS~\cite{Billinghurst_prab16}, steps in $f_\mathrm{inst,th}/f_\mathrm{s}$ as a function of $\sigma_\mathrm{z,0}$ can be seen around integer multiples. 
A connection to the number of substructures in the phase space is discussed amongst others in~\cite{Brosi_thesis_2020,Billinghurst_prab16}. 
In VFP solver simulations based on a pure parallel plates impedance model (see lower panel of Fig.~\ref{fig:finger}) the steps are less pronounced which is discussed in~\cite{Kuske_ipac17}. 

\begin{figure}
	\centering
	\includegraphics[width=1\columnwidth]{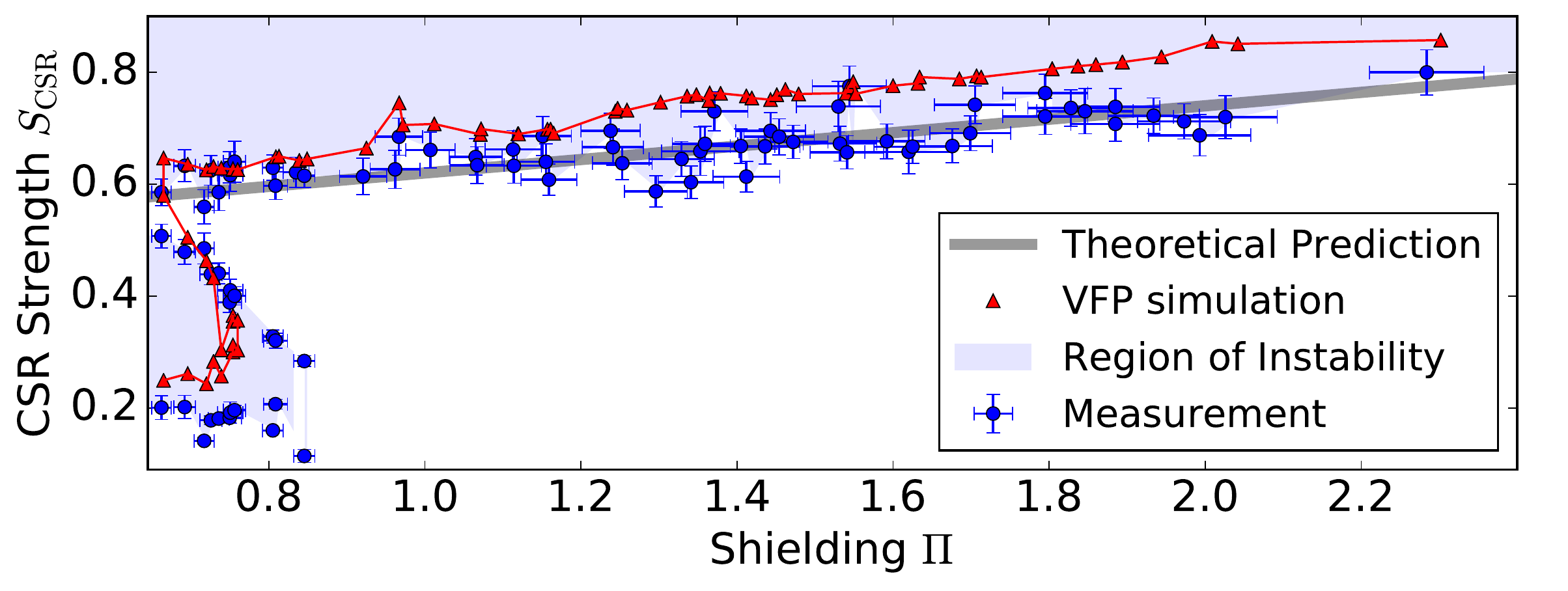}
	\vspace{-0.6cm}
	\caption{Measurements and simulations of the instability thresholds as well as the prediction by~\cite{Bane_cai_stupakov2010} are shown using the dimensionless parameter $\Pi$ and $S_\mathrm{CSR}$. The measured thresholds agree with the prediction while lying slightly lower than the simulations. Both show the additional region of instability (weak instability)~\cite{Bane_cai_stupakov2010,brosi_prab19} at lower values of $\Pi$ ($\approx$ shorter natural bunch length). Modified from~\cite{brosi_prab19}.}
	\label{fig:threshold}
			\vspace{-0.35cm}
\end{figure}

The most important characteristic of the micro-bunching instability is the instability threshold as the knowledge thereof is intrumental in avoiding operation in the instability.  
Figure~\ref{fig:threshold} shows the threshold current for different operational settings displayed in the dimensionless parameters\footnote{$\Pi= \frac{\sigma_{\mathrm{z},0}\, \rho^{1/2}}{ h^{3/2}}$, $S_{\mathrm{CSR}}= \frac{I_{\mathrm{n}}\, \rho^{1/3} }{ \sigma_{\mathrm{z},0}^{4/3}}$ with $I_{\mathrm{n}}=\frac{\sigma_{\mathrm{z},0}I_{\mathrm{b}}}{\alpha_{\mathrm{c}}\gamma\sigma_{\delta}^{2}I_{\mathrm{A}}}$, see~\cite{brosi_prab19}} $\Pi$ and $S_\mathrm{CSR}$. 
The measured thresholds coincide with the predicted simple, linear scaling law of the threshold $\left(S_\mathrm{CSR}\right)_\mathrm{th} = 0.5 + 0.12\,\Pi$~\cite{Bane_cai_stupakov2010} very well, while the simulated thresholds are slightly higher~\cite{brosi_prab16}. 
This is interesting, as the original simulations on which the prediction was based also lie slightly higher than the linear prediction within the measured region ($\Pi<2.3$). 
In general, it is not surprising that a pure parallel plates model does not completely describe the CSR interaction for KARA. 
This simple model does not consider any additional effects such as resistive wall, geometric impedances and edge radiation.
In \cite{schoenfeldt_diss_18}, for example, it was shown that an additional geometric impedance for an aperture slightly reduces the threshold current of the micro-bunching instability.
Nevertheless, as an estimation for the expected threshold current, the simple scaling law fits very well to these measurements. 

\vspace{-0.25cm}
\subsubsection{Multi-Bunch Studies}
\begin{figure}
	\centering
	\includegraphics[width=1\columnwidth]{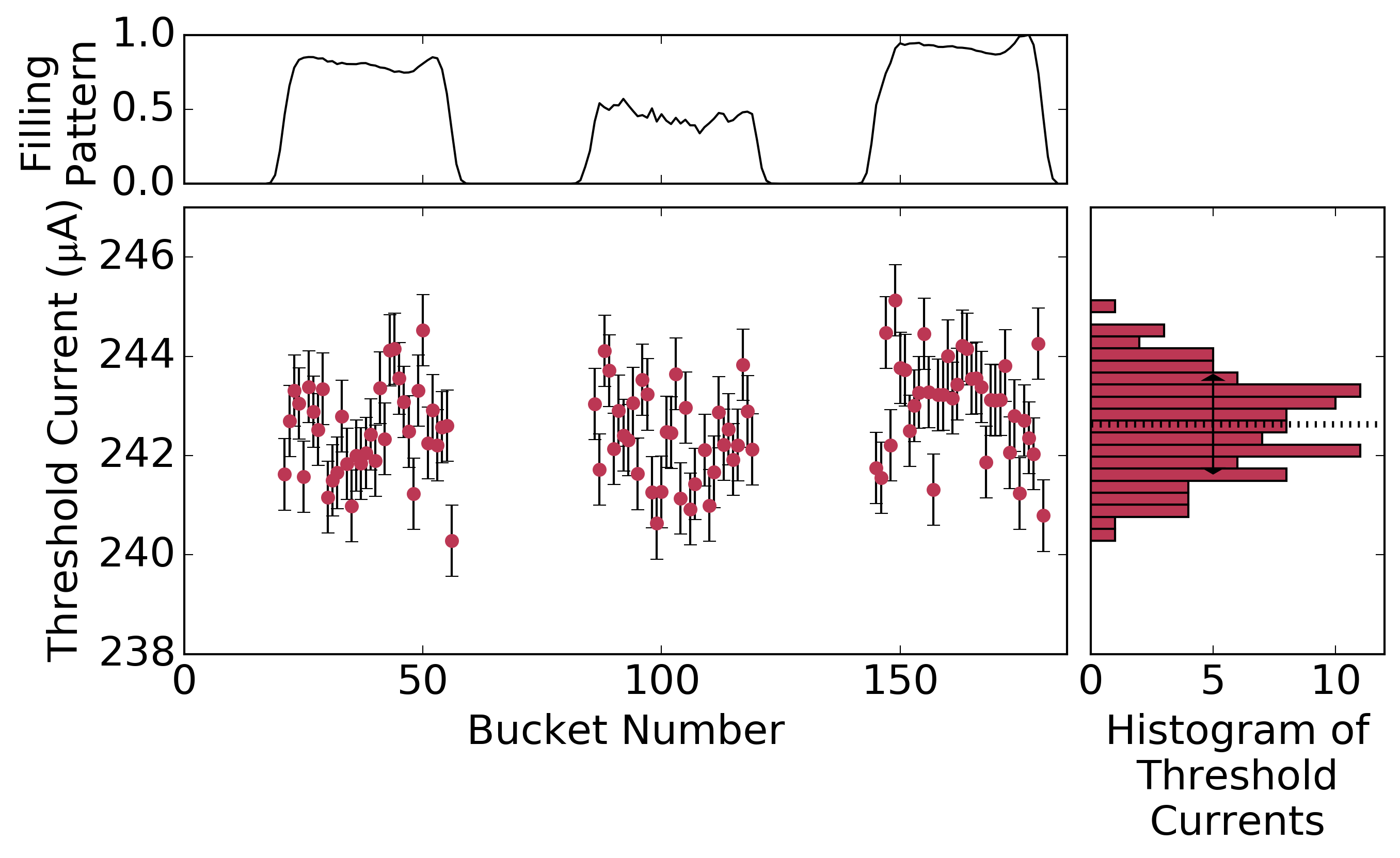}
	\vspace{-0.6cm}
	\caption{Measured threshold current of each bunch in multi-bunch operation with three bunch trains. Data from~\cite{Brosi_thesis_2020}. 
	}
	\label{fig:multi}
			\vspace{-0.45cm}
\end{figure}

As the DAQ system KAPTURE is capable of monitoring the emitted CSR power for each bunch during multi-bunch operation, it is possible to compare the behavior of the individual bunches and therefore study the influence of a multi-bunch environment on the micro-bunching instability. 
First indications for multi-bunch effects at KARA were shown in~\cite{anke2004,anke2012}. Moreover, possible effects of CSR based on whispering gallery modes were theoretically discussed and simulated~\cite{warnock_2013}.
Figure~\ref{fig:multi} shows the threshold current $I_\mathrm{th}$ measured for each bunch in a multi-bunch fill. 
The standard deviation of the distribution of the threshold currents is $\sigma\left(I_\mathrm{th}\right)=\SI{0.98}{\micro A}$. 
With an uncertainty on bunch current measurements of $\sigma_{I_\mathrm{b}\mathrm{,th}} =\SI{0.72}{\micro A}$ the remaining difference is $\sqrt{\sigma\left(I_\mathrm{th}\right)^2 - \sigma_{I_\mathrm{b}\mathrm{,th}}^2} = \SI{0.66}{\micro A}$.
Similar differences in the threshold currents in the order of less than \SI{1}{\micro A} were found for measurements at different operational parameters~\cite{brosi_ipac17}. 
Related studies were conducted for the instability frequency directly above the threshold~\cite{Brosi_thesis_2020}.
Investigations with a further improved bunch current resolution could yield interesting results.
Nevertheless, in all measurements the observed differences were small compared to the changes caused by changing the operational parameters.

\vspace{-0.2cm}
\subsection{Snapshot Method}
\begin{figure}[t] 
	\centering
	\subfloat[]{
	\includegraphics[width=0.93\columnwidth,trim={0 0.7cm 0 0}, clip]{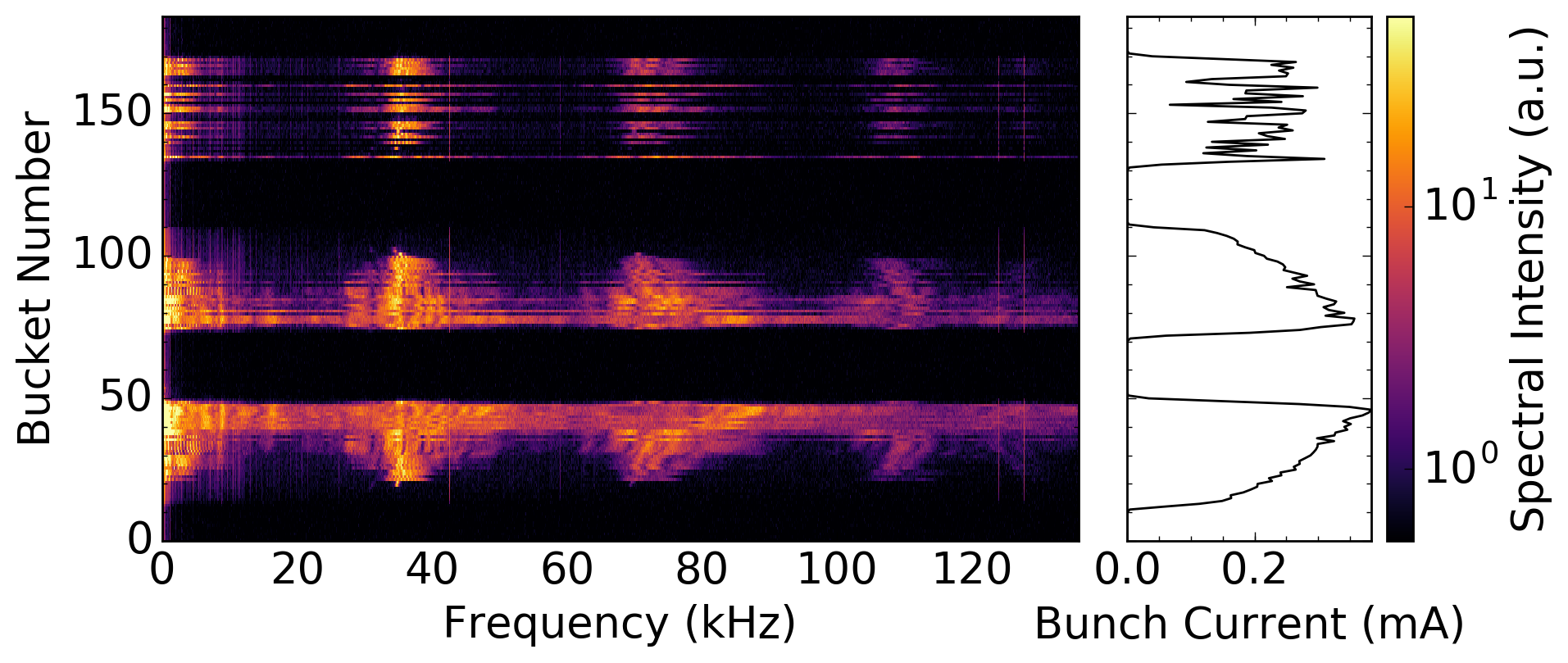}\label{fig:fft}
	}%
	\\
	\vspace{-0.45cm}
	\subfloat[]{
	\includegraphics[width=1.\columnwidth, trim={0 0.7cm 0 0},clip]{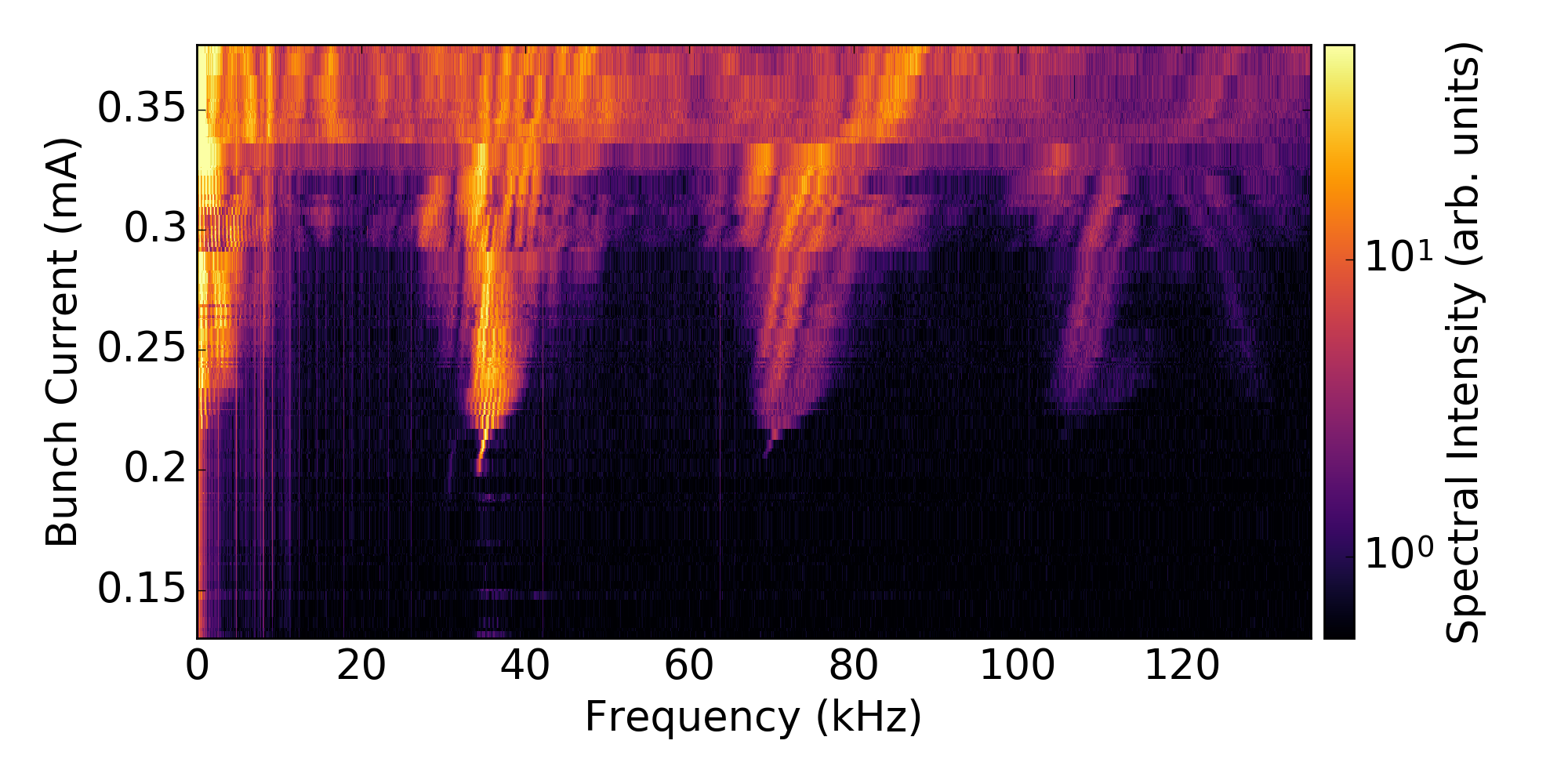}\label{fig:snap}
      }%
	\vspace{-0.15cm}
	\caption{(a) FFT of measured CSR power for each bunch in a multi-bunch fill. (b) Snapshot spectrogram obtained by sorting (a) by bunch current. The whole measurement took one second to provide a high frequency resolution.~\cite{brosi_prab16}}\label{fig:blub}
			\vspace{-0.55cm}
\end{figure}

With the result that the influence of a multi-bunch environment is small on the micro-bunching instability and the capability of KAPTURE to measure bunch-by-bunch, it was possible to develop a new measurement method. 
The snapshot measurement method reduces the measurement duration required for the recording of a spectrogram from hours to one second~\cite{brosi_prab16}. 
This is done by operating with a custom filling pattern and recording the CSR emitted by all bunches simultaneously for one second. 
Then the detector signal of each bunch is Fourier-transformed individually resulting in a plot as shown in Fig.~\ref{fig:fft}.
Sorting the FFTs of each bunch by their corresponding current results in a spectrogram as shown in Fig.~\ref{fig:snap}. 
The filling pattern is tailored so that all bunches have different bunch currents covering the whole bunch current range of interest. 
The resulting snapshot spectrogram clearly shows the same characteristic features as the single-bunch measurement shown in Fig.~\ref{fig:spec}, which was taken with the same operational settings. 
While the snapshot spectrogram has a limited current resolution, due to the limited number of bunches, it only took one second whereas the single-bunch measurement took 1.5 hours. 

The snapshot measurement method allows the fast characterization of the micro-bunching instability for the present operational settings, like a snapshot. 
This is extremely helpful for systematic studies, where now fast scans of different parameters and settings can be performed enabling a detailed comparison with simulations and fostering a better understanding. 
Furthermore, with regards to influencing and controlling the micro-bunching instability the snapshot method comes in very handy. 

\vspace{-0.2cm}
\section{ongoing work}
\vspace{-0.1cm}
The micro-bunching instability is still one of the main limitations for high current ($\gtrsim I_\mathrm{th}$) operation at short bunch lengths without reduction of the beam quality. 
So the design and development of a feedback system is a crucial step. 
A collaboration of PhLam and Soleil is working on a feedback system based on Pyragas time delayed feedback control~\cite{soleil_control_nature_2019}.
A further feedback system is being developed at KIT based on reinforcement learning and the capabilities of KAPTURE~\cite{boltz_icalepcs2019}. 
Both feedbacks rely on the RF system to interact with the bunch.
Another study focuses on the influence of additional impedances on the instability. 
In a collaboration of PhLam, Soleil and KIT an impedance chamber is being designed which will allow the insertion of additional impedances, e.g. a corrugated pipe structure, into the vacuum chamber~\cite{Maier_ipac21}.
A slightly different approach in~\cite{Boltz_ipac21} studies the possibility to further excite or even change the substructures by employing amplitude modulations of the acceleration voltage.
This has the potential to tailor the emitted CSR radiation and its fluctuations for possible applications of the THz radiation. 

Furthermore, the study of the micro-bunching instability is extended to other operation modes. 
A negative momentum compaction operation mode was established at KARA for investigations with regards to the applicability to future ultra-low-emittance machines~\cite{schreiber_ipac19}.
Also optics with low momentum compaction factor were established at the injection energy at KARA, to study the influence of the instability on the injection and vice versa~\cite{Papash_ipac19}. 

Steps towards the utilization of CSR emitted during the instability in form of steady-state micro-bunching are taken by a collaboration of the Tsinghua University Bejing, HZB, PTB and the Shanghai Light Source~\cite{SSMB_nature_2021,Feikes_ipac21}.

\vspace{-0.2cm}
\section{Conclusion}
\vspace{-0.1cm}
The micro-bunching instability is a longitudinal, collective instability which occurs in electron storage rings during the operation with short bunches. 
The complex dynamics during the instability lead to fluctuations in the emitted CSR as well as in the bunch length and energy spread and therefore also the horizontal bunch size. 
This instability was observed at many machines around the world and studied in a multitude of measurements to better understand the many dependencies on operational parameters. 
Dedicated diagnostics were developed which provide an even more detailed insight into the dynamics and reduce the necessary measurement durations significantly. 
The development of feedback systems to control the instability is underway and several ongoing studies focus on influencing or even driving the micro-bunching to make use of the increased emission of coherent synchrotron radiation. 

\vspace{-0.2cm}
\section{ACKNOWLEDGEMENTS}
\vspace{-0.1cm}
I want to acknowledge the fruitful discussions and cooperation with my dear colleagues at KIT, 
in particular, P.~Schreiber, J.~Steinmann, T.~Boltz, P.~Schönfeldt, B.~Kehrer, M.~Schwarz, J.~Gethmann, E.~Blomley and M.~Schuh for in-depth discussions on the dynamics as well as operation in special operation modes. 
My thanks go to M.~Caselle, L.~Rota, M.~M.~Patil, G.~Niehues, S.~Funkner, M.~J.~Nasse for the collaboration on DAQ systems and diagnostic tools. 
I greatly appreciate the support of Y.-L.~Mathis, B.~Gasharova, D.~A.~Moss and M.~Bank during measurements at the IR beamlines. 
Also, thanks to P.~Kuske (HZB) for the exchange about comparisons of simulations and measurements.
Last but not least, I want to thank R.~Ruprecht, E.~Bründermann and A.-S.~Müller for continuous support and advice allowing me to benefit from their broad experience. 

\vspace{-0.2cm}
\ifboolexpr{bool{jacowbiblatex}}%
	{\printbibliography
	\printbibitembibliography
	}%
	{%
	
} 
%
%


\end{document}